\documentclass[a4paper,11pt]{article}
\pdfoutput=1
\usepackage{jinstpub}
\usepackage{graphicx}
\usepackage{caption}
\usepackage{subcaption}
\usepackage{lineno}
\usepackage[font=scriptsize,labelfont=bf,skip=6pt]{caption}

\title{\boldmath Stability tests performed on the triple GEM detector built using commercially manufactured GEM foils in India}

\author[a,1]{M.~Gola, \note{Corresponding author.}}
\author[a,b]{S.~Malhotra,}
\author[a]{A.~Kumar,}
\author[a]{and~Md. Naimuddin}

\affiliation[a]{Department of Physics $\&$ Astrophysics\\ University of Delhi\\ Delhi, India}
\affiliation[b]{Now at Texas A\&M University\\ Doha, Qatar}

\emailAdd{mohit.gola@cern.ch}

\abstract{The Gas Electron Multiplier (GEM) technology is based on thin polymer foils cladded with copper on both the sides with a regular matrix of holes. Due to the limited manufacturing capacity of CERN, these GEM foils are now commercially manufactured also by Micropack Pvt. Ltd., a company based in India. In order to gain an insight on the behaviour of detectors built using the foils from Micropack, it is important to study various long and short term effects on these foils due to the applied voltage as well as the flux of the incident particles. In this paper, we report the effect on gain stability of  triple GEM detectors due to the polarising field induced by X-rays on the polyimide foils. Also, reducing the size of the amplifying structure to the microscopic scale results in a quick mitigation of the space charge effects which in turn helps in attaining a stable gain at very high incident flux. We report on the measurements of variations in the effective gain at very high particle flux of the order of $\rm MHz/mm^{2}$. }


\begin{document}
\maketitle
\flushbottom
\section{Introduction}
 
The Gas Electron Multiplier (GEM) was introduced by Fabio Sauli, who was working at European Center for Nuclear Research (CERN) in 1997 \cite{Sauli, Sauli2, Sauli3, Hoch} and since then it has been used for various practical applications. Currently, many high energy and nuclear physics experiments are using or proposing to use GEM technology, which is creating a big demand for GEM foils. Therefore it is difficult for CERN, which is the main provider of GEM foils, to keep up with such an increasing demand. There is a need for commercially available GEM foils to help fulfill the surge in demand. The Micropack Pvt. Ltd., an India based company, acquired a license from CERN under Transfer of Technology (ToT) to produce GEM foils. The University of Delhi (DU) then began collaborating with Micropack to help them establish a consistent manufacturing procedure for GEM foils utilizing both double-mask and single-mask techniques \cite{MerlinThesis, DblMask}. The Micropack $\rm 100~mm \times 100~mm$ double-mask GEM foils that have been optically analyzed for geometric properties and electrically tested via leakage current at DU have been discussed elsewhere~\cite{Foils}. The basic quality controls on the detector built using these foils are described in~\cite{PerfDU}.

In this paper, we will present studies performed on a detector built using commercially manufacture GEM foils. These studies are helpful to understand the characteristic of the insulating material (Apical \footnote[1]{Apical is a polyimide film similar to Kapton and is available from Kaneka Texas Corporation, Pasadena, TX.}) and how they affect the stability of GEM detector. The gain stability \cite{Pol_Volt, Stab_GEM_TPC} over time is a very important parameter of gaseous detectors because any unwanted variation in gain can cause a loss of efficiency.  The rate capability \cite{EveRC} of a gaseous detector shows the variation of detector gain with respect to the incident flux and for Micro Pattern Gaseous Detector (MPGD) the gain stays stable up to very high incoming rates but it also depends upon space charge effects \cite{Space_charge}. Several systematic studies on such effects have been performed in the past to determine the principle causes of the gain variation observed in the detector due to the properties of foils. We will report on some of these studies here.\\ 

In the following section, we describe the experimental set-up used for performing the studies mentioned in this paper. Section 3 is dedicated to understanding the gain stability of the detector whereas section 4 provides a quantitative idea of the behaviour of the triple GEM detector under very high incoming particle flux which is $O(MHz/mm^2)$. Finally, we conclude with discussions on the results obtained.

\section{Experimental set-up for measurements}
\begin{figure}[h]
        \hspace{1 cm}
    \begin{subfigure}[b]{0.25\textwidth}
    \centering
    \includegraphics[width=\textwidth]{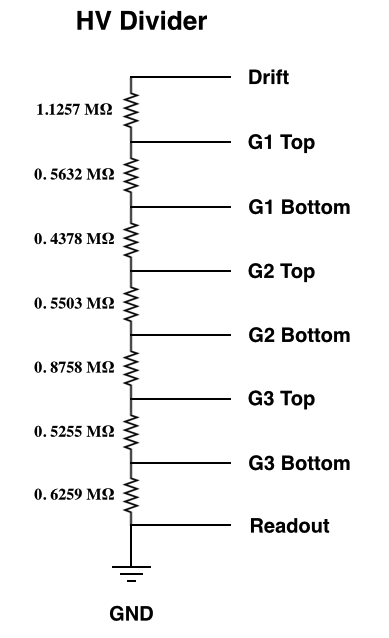}
    \caption{}
    \end{subfigure}
    \hspace{2.5 cm}
    \begin{subfigure}[b]{0.4\textwidth}
    \centering 
    \includegraphics[width=\textwidth]{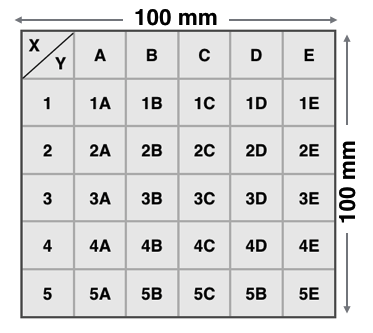}
    \caption{}
    \end{subfigure}
    \caption{Schematic of (a) high voltage resistive divider used for triple GEM detector prototype, and (b) Nomenclature used for 5 $\times$ 5.} 
\label{fig:Grid}
\end{figure}

The active area of the GEM foils used for this study is $\rm 100~mm \times 100~mm$; they have been manufactured with the double-mask etching technique to have 70 $\rm\mu m$ (50 $\rm \mu m$) outer (inner) hole diameter and pitch of 140 \rm $\mu m$. The triple GEM detector was built with 3mm/1mm/2mm/1mm for Drift/Transfer1/Transfer2/Induction gap configuration and all the foils were powered up using a High Voltage (HV) resistive divider whose schematics is shown in fig.~\ref{fig:Grid}(a). The active area of the GEM detector was divided into 5 $\times$ 5 equal sectors each covering an area of $\rm 20~mm \times 20~mm$ and their nomenclature was fixed as shown in fig. \ref{fig:Grid}(b).  A 1 mm collimator was placed in the front of the X-Ray and it was placed perpendicular to the detector touching the drift volume. Therefore, while exposing the particular sector with X-Rays, our setup ensured that no other neighboring sectors got exposed to the beam.

The gas mixture used in all the measurements was a mixture of noble gas and a quencher \cite{GasOperation}. For the studies reported here, a mixture of Argon ($\rm Ar$) and Carbon-Dioxide ($\rm CO_2$) in the ratio of 70\% and 30\% respectively have been used. The main advantage of those mixtures is that these gases are non-flammable, eco-friendly and that they are easily available and relatively cheap \cite{GasProp,Gases}. A gas mixing unit is used for supplying the fixed mixture and the flow rate was controlled by the mass flow controllers which has been set to 5 $\ell$/h for all the measurements.

The GEM detector has been powered up using a CAEN N1470 multichannel power supply. A high voltage filter is introduced between the power supply and the detector to reduce the electronics noise. The output is read by the readout board having an active area of $\rm 100~mm \times 100~mm$ and consisting of 128 1-D strips. Signal amplification was performed in two stages; first it was amplified by ORTEC 142 charge sensitive pre-amplifier and then using the ORTEC 474 timing filter amplifier. For counting measurements, the signal passed through ORTEC 974 discriminator with a threshold of 140 $\rm mV$ to cut the noise. The rate has been measured using CAEN N1145 Scaler \& Counter over the desired period of time. The current was measured with the help of Keithley 6517B pico-ammeter \cite{PicoAmp} and recorded with a Labview program via a GPIB interface. 


\section{Gain stability}

\begin{figure}[h]
    \begin{subfigure}[b]{0.5\textwidth}
    \centering
        \includegraphics[width=\textwidth]{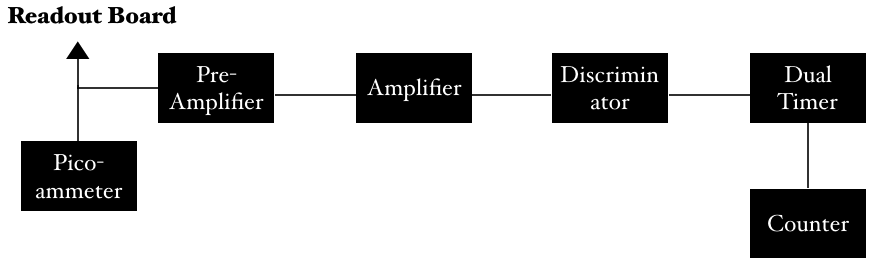}\qquad
        \caption{}
    \end{subfigure}
  \hspace{0.5cm}
    \begin{subfigure}[b]{0.42\textwidth}
     \centering
        \includegraphics[width=\textwidth]{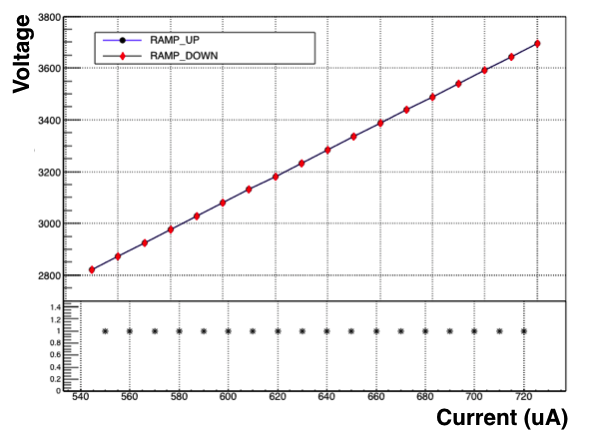}
        \caption{}
    \end{subfigure}
\caption{(a) Effective gain measurement block diagram.(b) I-V Characteristics of the detector showing ohmic behaviour while ramping up (black circles) and ramping down (red diamonds) at different operating voltages(right).}
\label{fig:IV_UP_DOWN}
\end{figure}

The effective gain of a detector is a unique parameter used to relate the general properties which include the electrical and geometrical properties together with the gas composition. This is defined as the ratio of the charge arriving at the anode and the charge created in the drift volume.
  \begin{center} $G = \frac{I_{with~Source} - I_{Without~Source}}{No.~of~Primaries*rate~on~Anode*charge~of~electron}$ \end{center}
Figure~\ref{fig:IV_UP_DOWN}(a) shows the block diagram of how the current and rate was measured from the detector.  The gain stability is very important for gaseous detectors because any unwanted variation in gain causes a change in efficiency. As mentioned high voltage divider is used to power up the GEM foils and using it a current-voltage (IV) curve is obtained when the voltage is first ramped up and then ramped down across the divider. The measured current shows slope for ramping up and ramping down and hence no hysteresis in IV has been observed \cite{Hysteresis_IV}.  Figure~\ref{fig:IV_UP_DOWN}(b) shows an ohmic behaviour at a given range of divider current across the detector with a total resistance of the circuit as 5 $\rm M\Omega$ (which includes the resistance of HV filter as 0.3 $\rm M\Omega$ and HV divider as 4.7 $\rm M\Omega$).

\begin{figure}
\centering
\includegraphics[width=0.5\textwidth]{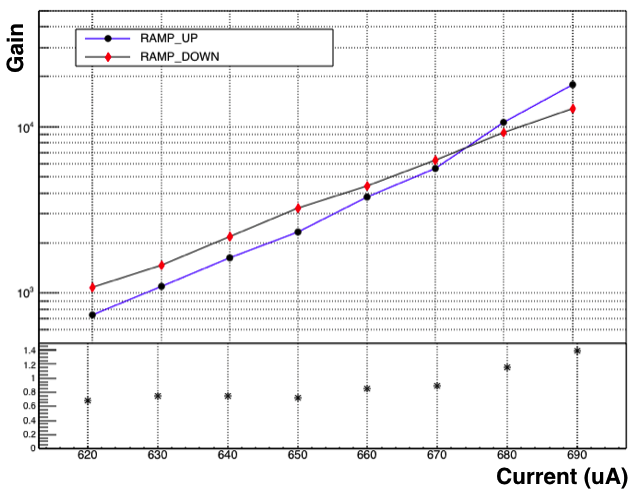}
\caption{Effective gain of the detector as a function of current across the detector while ramping up (black circles) and ramping down (red diamonds) with Ar/$\rm CO_{2}$ = 70/30.}
\label{fig:70_30_UP_DOWN}
\end{figure}

In order to understand the hysteresis as a function of other parameters of the GEM detector, a measurement of effective gain has been performed. The measurement has been done while ramping up and then ramping down the divider current across the detector. The outcome of this exercise is shown in Figure~\ref{fig:70_30_UP_DOWN}. The measurements were taken by varying the current in steps of 10 $\rm \mu A$ from 620 $\rm \mu A$ to 690 $\rm \mu A$ by setting the corresponding voltage and then in the reverse direction with waiting time of 2 minutes between each measurement. As expected, the gain varies exponentially with respect to the current. However, looking at the variation in the log scale, the slope of the gain has been found to be different while ramping up and ramping down. This is a manifestation of hysteresis effect in gain \cite{Hysteresis_Gain} of the detector which is due to the charging up or polarisation effects in the detector/foils.

The gain stability of a triple GEM detector is affected mainly by two effects: the charging up and the polarisation effects. The charging up effect occurs due to the multiplication of the charge near the surface of the foil which results in the trapping of the charge in the hole~\cite{IEEE} . The polarisation effect, on the other hand, is due to the movement of the charges inside the polyimide layer after applying the voltage across the foil. It is independent of the charge deposited by the particle but depends upon the geometrical and electrical properties of the foil. The polarisation effect can be explained with the help of a trapping model, as described in~\cite{Pol_Trap} for Kapton-H film. Due to the absorption of the photon by polyimide during the irradiation of X-rays, electron excitation in the polyimide occurs. The presence of an electric field across the foil makes these electrons move in its direction unless they get captured by some trapping center in the foil. This process results in an increase of the anode current which in turn increases the gain. However, when the dynamic equilibrium is reached the gain starts to saturate.\\

To estimate the gain stability in time and variations in the gain due to polarisation effect, a series of measurements of gain of the detector was carried out for several hours. The anode current was recorded while irradiating a sector of the detector with Amptek Mini-X X-ray \cite{Amptek} source for 30 seconds. This process was repeated at an interval of 5 minutes for the same sector until a stable gain value had been achieved. These measurements were performed at various values of gain of the detector to estimate the dependency of divider current (or gain) on polarisation effect. During the measurement ambient temperature was continuously monitored using an ARDUINO based system.\\

\begin{figure}[h]
\centering
\includegraphics[width=0.45\textwidth]{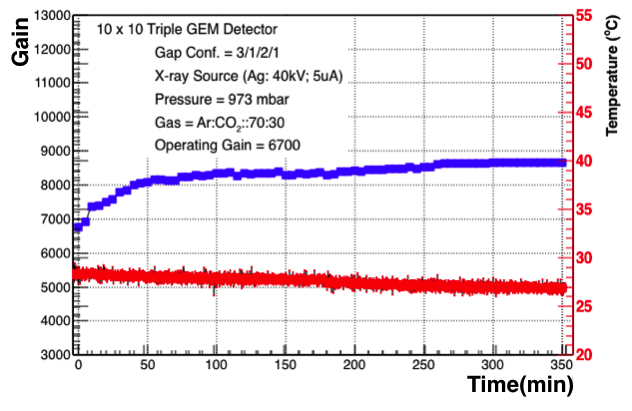}
\caption{Variation of effective gain (blue squares) of the detector as a function of time having initial gain of 6,800 for triple GEM detector in Ar/$\rm CO_{2}$ = 70/30. Variation of temperature was also recorded every second and plotted (red continuous line).}
\label{fig:Pol_15k}
\end{figure}

\begin{figure}[h]
    \begin{subfigure}[b]{0.455\textwidth}
    \centering
        \includegraphics[width=\textwidth]{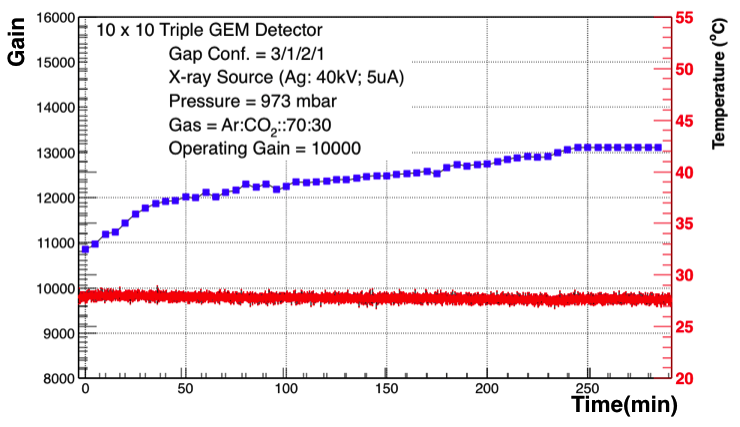}\qquad
        \caption{}
    \end{subfigure}
  \hspace{0.5cm}
    \begin{subfigure}[b]{0.42\textwidth}
     \centering
        \includegraphics[width=\textwidth]{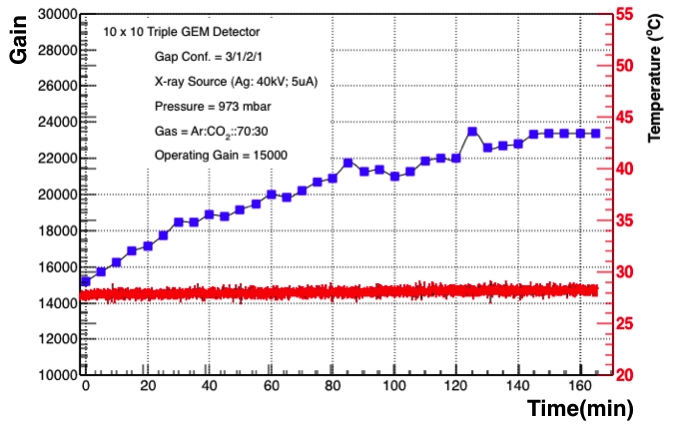}
        \caption{}
    \end{subfigure}
\caption{Variation of effective gain (blue squares) of the detector as a function of time having initial gain of (a) 10,000, and (b) 15,000 for triple GEM detector in Ar/$\rm CO_{2}$ = 70/30. Variation of temperature was also recorded every second and plotted (red continuous line).}
\label{fig:Pol_10k_6k}
\end{figure}

Figure~\ref{fig:Pol_15k} shows the value of gain with respect to time. The measurement started with a gain of 6.8k and the plateau was observed after 6 hours showing a difference of $\sim$29\% in gain. During the measurement, the ambient temperature was monitored continuously.  From the measurement it is visible that the percentage difference of change in gain and the time taken to reach the stable gain depends upon the initial value of the gain of the detector. For an initial gain of 11k and 15k, the time to reach the gain plateau is 5 hours and 2.5 hours respectively as shown in Figure~\ref{fig:Pol_10k_6k}. From this measurement, we conclude that the gain variation due to the polarisation can be mitigated by switching ON the chamber for several hours prior to the start of precise measurements of timing and efficiency of the detector.

\begin{figure}[h]
\centering
\includegraphics[width=0.42\textwidth]{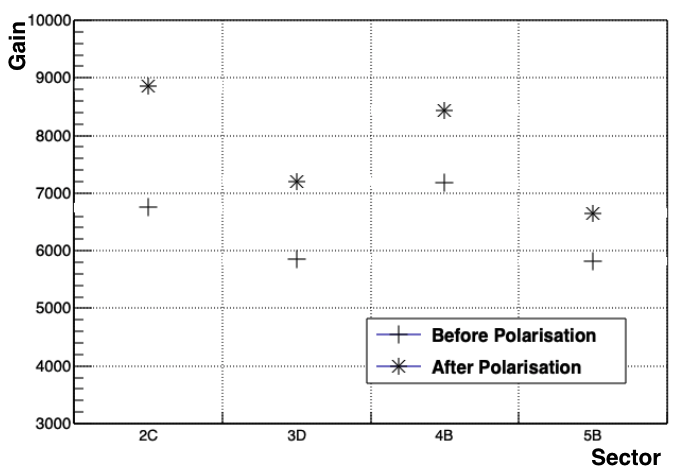}
\caption{Comparison between the initial and final gain scans for the polarisation measurement of the triple GEM detector. The detector was irradiated at position 2C by X-ray source with an initial gain of 6,751.}
\label{fig:Global}
\end{figure}

Since the polarisation effect is a global phenomena, a gain scan was performed before and after the test to disentangle the local and global fluctuations. A comparison of gain at different positions before and after the polarisation effect is shown in fig.  \ref{fig:Global}. Initially, the X-ray source was placed at sector 2C according to the nomenclature in section 2 for $\sim$6 hours. The gain at the position (3D, 4B,and 5B) which were not irradiated is increased by a few tens percent with respect to the initial gain value. And this increase in gain is due to the polarisation which is a global phenomena.


\section{Rate capability}

The MPGD technologies were mainly introduced in response to the limited rate capability of the Multi-wire Proportional Counters (MWPC) to handle fluxes higher than several $\rm kHz/mm^2$. Reducing the amplifying structure to the microscopic scale helps quickly mitigate the effect of space charge which results in higher gains even for the high incoming particle flux \cite{RC1}.

GEM detectors are known for their stable operation even at very high particle flux. In the particular case of the triple-GEM technology, we can distinguish three different regions depending on the incoming flux of particles. The detector shows stable gain when the incoming flux is of the order of a few tens of $\rm kHz/mm^2$ (horizontal region), at a particular value of divider current (or gain) across the detector. As the flux increases to few $\rm MHz/mm^2$ (upward region) the gain increases as well. Further increase in the value of flux results in the decrease of the gain (downward region). 

To check the dependency of the effective gain with the X-ray flux, known as rate capability, a collimated beam of 22.1 KeV X-rays of about 1 mm beam diameter from a silver X-ray generator has been used to produce the primary ionisation in the conversion volume.

\begin{figure}[!ht]
    \begin{subfigure}[b]{0.46\textwidth}
	\centering	
        \includegraphics[width=\textwidth]{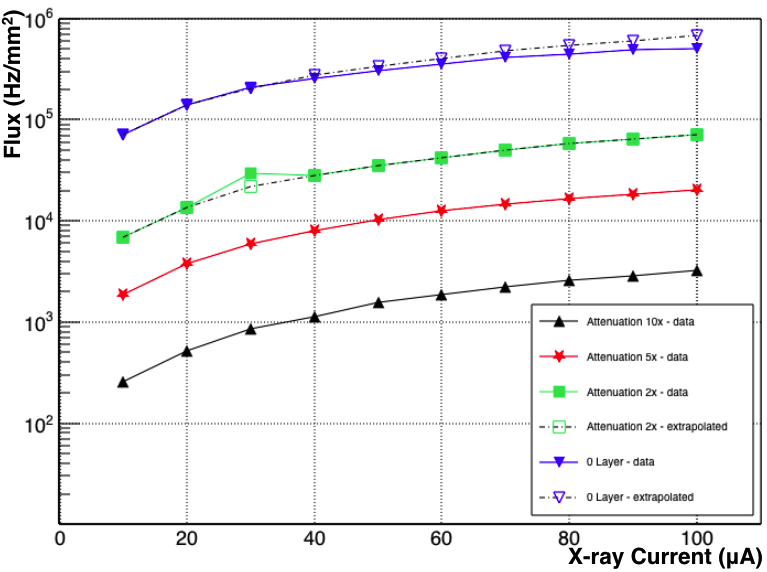}\qquad
        \caption{}
    \end{subfigure}
        \hspace{0.5cm}
    \begin{subfigure}[b]{0.5\textwidth}
    \centering
        \includegraphics[width=\textwidth]{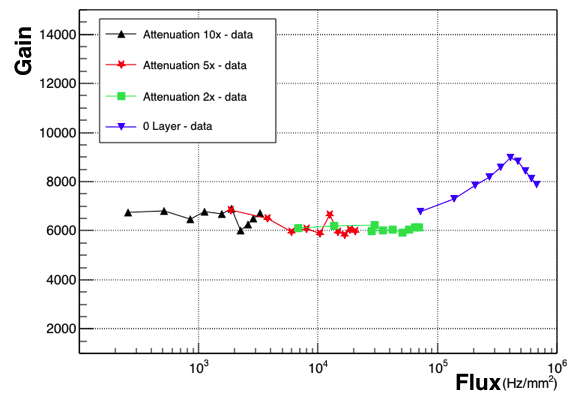}
        \caption{}
    \end{subfigure}
\caption{(a) Flux provided by the X-ray source using different layers of Copper attenuators vs. the X-ray source supply current, and (b) Rate capability for triple GEM detector operated at a nominal effective gain of approximately 6.5$\rm k$ in Ar/$\rm CO_{2}$ = 70/30.}
\label{fig:RC}
\end{figure}

To estimate the effective gain as a function of the particle flux, the X-ray was placed on a particular sector to measure the amplified detector current. The X-ray flux was adjusted by changing the X-ray tube power or by attenuators. The flux delivered to the detector was calculated by taking into account the X-ray rate measured by the discriminator and the known diameter of the collimator. Since, for the higher rate, discriminator starts to saturate due to pile up, the interaction rate on the detector was measured using the copper attenuators. Once the attenuation factor is known, the interaction rate can be extrapolated to obtain the rate without attenuator. Figure~\ref{fig:RC}(a) shows the estimated flux for different power of the X-ray source. Figure~\ref{fig:RC}(b) shows the measurement of rate capability at the initial detector gain of 6800. The detector gain is stable from lowest flux used up to about 50 $\rm kHz/mm^2$. For higher fluxes, up to approximately 0.4 $\rm MHz/mm^2$, the effective gain increases as a function of flux. Further increasing the flux results in a decrease of the effective gain.

\begin{figure}[!ht]
    \begin{subfigure}[b]{0.5\textwidth}
    \centering
        \includegraphics[width=\textwidth]{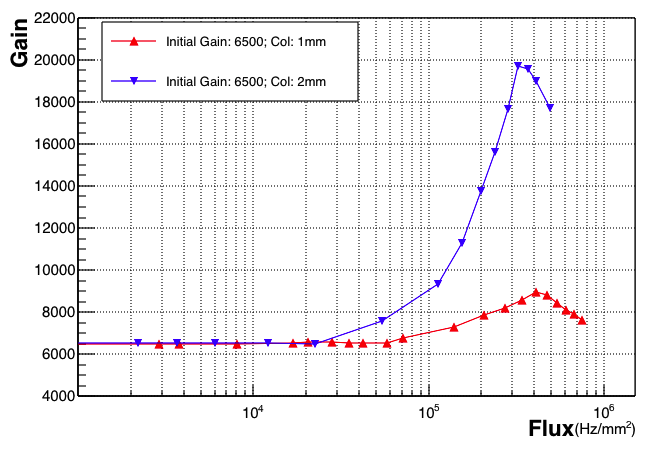}\qquad
        \caption{}
    \end{subfigure}
    \hspace{0.5cm}
    \begin{subfigure}[b]{0.5\textwidth}
    \centering
        \includegraphics[width=\textwidth]{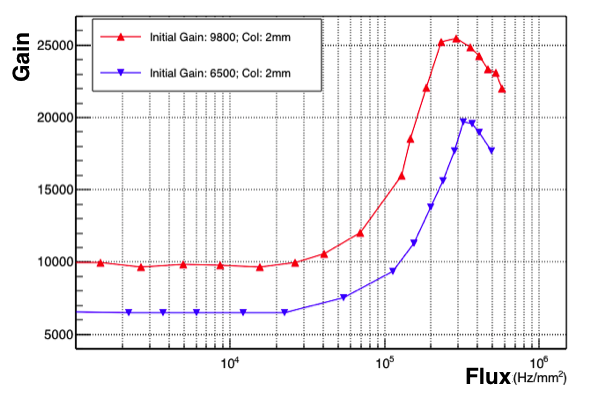}
        \caption{}
    \end{subfigure}
\caption{Dependence of effective gain as a function of flux (a) having same initial gain but different collimator, and (b) different initial gain but same collimator for triple GEM detector in Ar/$\rm CO_{2}$ = 70/30.}
\label{fig:Comb_RC}
\end{figure}

The rate capability studies were further extended by using two different collimators in front of the X-ray source, and the result is shown in Figure~\ref{fig:Comb_RC}(a) with an initial detector gain of 6,500 for both collimators. By changing the collimator from 1 $\rm mm$ to 2 $\rm mm$, the area of the sector under irradiation increases 4 times which causes a steeper increase in the effective gain. The rate capability measurement was performed for initial gains of 6500 and 9800 with 2 mm collimator, as shown in Figure \ref{fig:Comb_RC}(b). The observed change for different initial gain is related to the charge density in the detector. And higher nominal gain leads to the appearance of the observed effects at lower particle fluxes with the increase being steeper while a decreased gain will shift the effect towards higher fluxes. This is due to the fact that the field distortion depends upon the number of ions generated and accumulated in the detector. For a higher nominal effective gain of 9800, the number of ions in the detector volume is larger than compared to a detector operated at a gain of 6500.


\section{Summary}
A study has been carried out to investigate the gain stability and rate capability of the triple GEM detector assembled using commercially manufactured GEM foils. During the high voltage scan of the detector, no hysteresis was observed while a significant effect was observed in the effective gain measurement during ramping up and down the divider current. The short term stability of the GEM foil was measured and a direct correlation has been established with the initial gain of the triple GEM detector. For an initial gain of 6.8K, the plateau observed in 6 hr was found to have a percentage difference of $\sim$29\% with respect to the initial gain. However, increasing the divider current sees an increase in the initial gain as the plateau effect has been observed much earlier. Also, the measurement of rate capability of the triple GEM detector used shows stable gain up to flux of about 50 $\rm kHz/mm^2$ with 1 mm collimator. A correlation between the rate capability with respect to the different initial gain as well as different collimators placed in front of X-ray has also been performed and results compared. Using a collimator of relatively bigger diameter causes a stepper increase in the gain and varying the initial gain results in the occurrence of the effect earlier due to charge density.
 
 
\section{Conclusions}
Micropack has been successful in manufacturing the small-area Double mask GEM foils. However, before these foils can be utilised for various applications, it is important to characterize these foils and the detectors assembled from them for various properties. In this study, we tested these foils for their short term stability and the rate capability. 
The polarization and rate capability of detectors obtained using commercially manufactured foils in India is similar to the detectors built with CERN made foils \cite{RC_Comp}. Therefore, these foils can be used for various purposes and the technology can be exploited in interdisciplinary applications such as medical, cultural heritage studies, muography, etc. Currently, Micropack Pvt. Ltd. is one of the candidates to build the foils for the smallest modules of GE2/1 size detectors as well as for the ME0 \cite{TDR} CMS muon chamber upgrade. The studies reported in this manuscript are a step in the direction of utilising these foils in the CMS experiment and elsewhere.


\section{Acknowledgements}
We would like to acknowledge the funding agency, Department of Science and Technology (DST), New Delhi (grant nos. SR/MF/PS-02/2014-DUA(G) and SB/FTP/PS-165/2013) for providing financial support. 


\end{document}